\documentclass[twocolumn,letter]{jpsj2} 
\title{Spin dynamics in copper metaborate CuB$_2$O$_{4}$ studied by muon 
spin relaxation}

\author{\textsc{A.~Fukaya}$^{1}$\thanks{Present address: RIKEN (The Institute of Physical and Chemical Research), Wako, Saitama 351-0198, Japan, E-mail address: afukaya@riken.jp}, \textsc{I.~Watanabe}$^{2}$, \textsc{K.~Yamada}$^{1}$ and \textsc{K.~Nagamine}$^{3}$}

\inst{$^{1}$Institute for Materials Research, Tohoku University,
Katahira Aobaku, Sendai 980-8577, Japan\\
$^{2}$RIKEN (The Institute of Physical and Chemical Research), Wako, Saitama 351-0198, Japan\\
$^{3}$Department of Physics, University of California, Riverside, 
  Riverside, CA 92521, U.S.A}

\abst{Copper metaborate CuB$_2$O$_{4}$ was studied by muon spin 
relaxation measurements in order to clarify its static and dynamic 
magnetic properties. The time spectra of muon spin depolarization suggest that the local fields at the muon site contain both static and fluctuating components in all ordered phases down to 0.3~K. In the weak ferromagnetic 
phase (20~K~$>T>$~9.3~K), the static component is dominant. 
On the other hand, upon cooling the fluctuating component becomes dominant 
in the incommensurate helix phase (9.3~K~$>T>$~1.4~K). The dynamical 
fluctuations of the local fields persist down to 0.3~K, where a new 
incommensurate phase ($T<$~1.4~K) is expected to appear. This result 
suggests that spins fluctuate even at $T$$\rightarrow$0. We propose two 
possible origins of the remnant dynamical spin fluctuations: frustration of the exchange interactions and the dynamic behavior of the soliton lattice.
}

\kword{copper metaborate, muon spin relaxation, spin dynamics, magnetic soliton lattice, weak ferromagnet, incommensurate state}

\begin{document}
\makeatletter
\def\@citess#1{\textsuperscript{#1)}}

\maketitle


Soliton lattice is an interesting structure where domain walls (solitons) with incommensurate state separate commensurate regions and are arranged periodically~\cite{Bak82}. The soliton lattice structure has rarely been found in pure localized spin systems. Only a few materials have been discovered; $e.g.$, spin-Peierls systems CuGeO$_3$~\cite{Boucher} and (TMTTF)$_2$PF$_6$~\cite{Brown} in magnetic fields, the two-dimensional spin system Ba$_2$CuGe$_2$O$_7$~\cite{Zheludev97,Zheludev99}, and the recently-discovered copper metaborate CuB$_2$O$_4$~\cite{Roessli01}. While the static properties of the magnetic soliton lattice have been studied experimentally, the dynamic properties have not been clarified. In this paper, we focus on the spin dynamics of CuB$_2$O$_4$ and report its magnetic properties. 

In CuB$_2$O$_4$, three magnetic phase transitions occur down to 50~mK~\cite{Petrakovskii02}. A second order phase transition from a paramagnetic phase to a commensurate non-collinear weak ferromagnetic phase occurs at $T_N$$\sim$21 K. A first order phase transition to an incommensurate helix phase occurs at $T^*$$\sim$10 K. At a narrow temperature region around $T^*$, higher-order magnetic satellites are observed by the neutron scattering measurements, suggesting the formation of a magnetic soliton lattice~\cite{Roessli01}. Recently, a third phase transition has been suggested to occur at $\sim$1.8 K, where a sudden change is observed in the intensity of the satellites~\cite{Petrakovskii02,Boehm02,Boehm03}. However, details of the phase transition and the magnetic state in this phase have not yet been clarified. Hereafter, we will refer to this low-temperature incommensurate phase as ``LI phase.''

NMR measurement, which is a powerful method to study spin dynamics, has been performed for CuB$_2$O$_4$.~\cite{Nakamura04,Nakamura05} However, the dynamic behavior in zero field is still unclear because of the quite large relaxation rate and necessity of applying magnetic fields in NMR method. 
The muon spin relaxation ($\mu$SR) method is another powerful technique to study both magnetic order and spin dynamics in magnetic fields and zero field~\cite{Schenck85}. In fact, Boehm $et$~$al.$ have already performed $\mu$SR measurements on CuB$_2$O$_4$. Interestingly, in contrast to the clear phase transitions observed in the magnetization, as shown in Fig.~1(a), they reported no muon-spin oscillation down to 30~mK~\cite{Boehm02}. However, some amplitude of the signal appears to oscillate in the weak ferromagnetic phase. In addition, Boehm $et$~$al.$ presented no information on the longitudinal field (LF) dependence of the time spectrum, which is indispensable for the study of spin dynamics. Therefore, we have performed detailed $\mu$SR measurements for CuB$_2$O$_4$. 
In $\mu$SR measurements, muon spin depolarization is measured as a function of time. In a isotropic powder sample, if local fields are static, the muon spin polarization at long time (so-called `` long-time tail'') persists at 1/3, since on average 1/3 of muon moment is parallel to local fields and do not contribute to the depolarization. On the other hand, if local fields are fluctuating, the polarization relaxes to zero. Therefore, we used a powder sample in order to easily determine whether the local fields are static or fluctuating.


A powder sample of CuB$_2$O$_4$ was prepared by solid state reactions. A platinum crucible was used for the reaction. When we mixed the starting materials, B$_2$O$_3$ and CuO, with stoichiometric ratio, the resulting substance hardened and could not be ground. In order to avoid this problem, we first mixed 1/3 of the stoichiometric amount of B$_2$O$_3$ with CuO, heated it in air at 950$^\circ$C for 12 hours, then ground it. This procedure was repeated two more times. Finally, in order to complete the synthesis, the substance in which B$_2$O$_3$ had completely been added was heated. An X-ray powder diffraction of the sample was qualitatively analyzed, and it indicated a single phase of CuB$_2$O$_4$. 

We measured the temperature dependence of the magnetization $M$ for characterization. The results are shown in Fig.~1(a). As the temperature decreases, $M$ increases at $T_{\rm N}=20.3$~K and decreases abruptly at $T^*=9.5$~K. These two temperatures are slightly lower than the results of Ref.~\citen{Petrakovskii99}. 

We performed $\mu$SR measurements using a pulsed muon beam at the RIKEN-RAL Muon Facility in the U.K. The temperature was controlled with a $^3$He cryostat in the temperature region of $0.3-70$~K. In this paper, we discuss the time spectra of muon spin depolarization by using the asymmetry of count rate between the forward and backward counters after subtracting the background. 


\begin{figure}[t]
\begin{center}
\includegraphics[width=6.7cm]{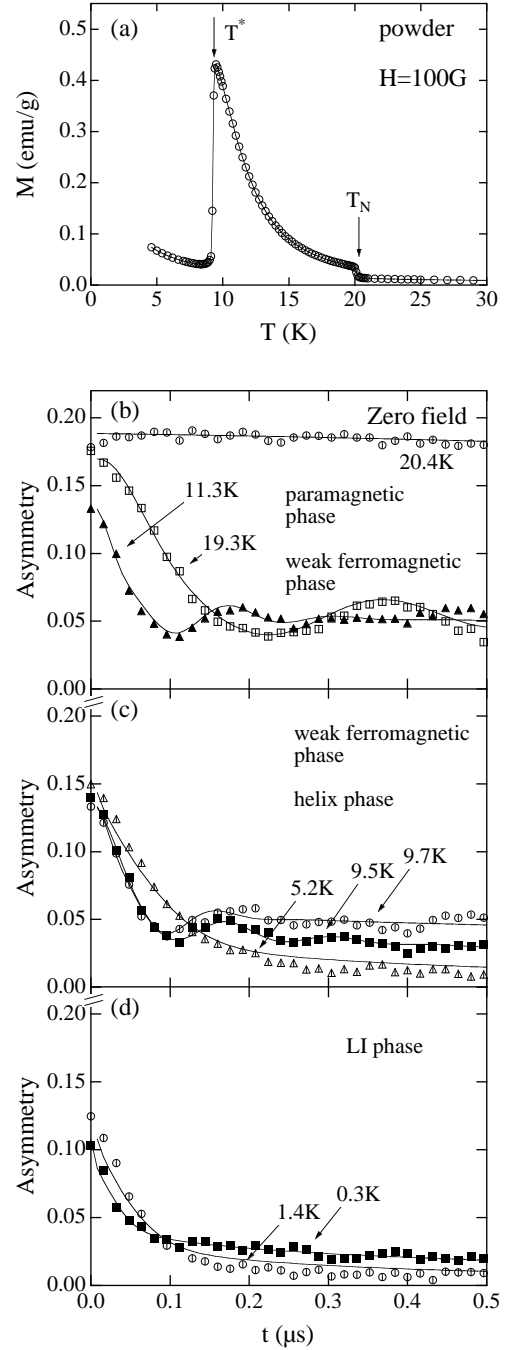}
\end{center}
\caption{(a) Magnetization of CuB$_2$O$_4$ as a function of temperature measured in a magnetic field of $H=100$~G. No significant irreversibility was observed in $M$ in the zero-field cooling, field-cooling, and field-heating conditions. $T_{\rm N}$ and $T^*$ are shown by the arrows. (b)$-$(d) $\mu$SR time spectra in zero field measured at (b) $T>T^*$, (c) $T = \sim$$T^*-5.2$~K, and (d) $T\leq1.4$~K. The solid lines show the results of the fitting.
}
\label{f1}
\end{figure}

The temperature dependence of the time spectrum in zero field (ZF) is shown in Fig.~1(b)$-$(d). The spectra in the early-time region of $t<0.5~\mu$s are shown. Above 20.4~K (paramagnetic phase), the relaxation is slow, which shows that the correlation time of the electronic spins is too short to contribute to the time spectrum. As the temperature decreases to $T_{\rm N}$, oscillations are clearly observed in the time spectrum, which shows the existence of a magnetic long-range order. This observation differs from the report in Ref.~\citen{Boehm02}. As shown in Fig.~1(b), the frequency of the oscillations increases with decreasing temperature. In the weak ferromagnetic phase, the asymmetry at $t \sim 0.2$~$\mu$s is approximately 1/3 of the full initial asymmetry (the initial asymmetry in the paramagnetic phase). Such time spectra are characteristic of static local fields in isotropic powder samples.

Interestingly, the oscillations disappear upon cooling below $T^*$, as shown in Fig.~1(c) and (d). The asymmetry is clearly lower than 1/3 of the full initial asymmetry at $t>0.2~{\rm \mu}$s. These results suggest that the fluctuating component becomes predominant below $T^*$. Upon cooling down to 5.2~K, the relaxation in the early-time region ($t< 0.1~\mu$s) becomes slower, and the asymmetry for $t> \sim$0.2~$\mu$s approaches zero. Upon more cooling, the relaxation becomes fast again, and the asymmetry $t>0.1~\mu$s recovers slightly at $\sim$0.3~K.  


\begin{figure}[t]
\begin{center}
\includegraphics[width=7.4cm]{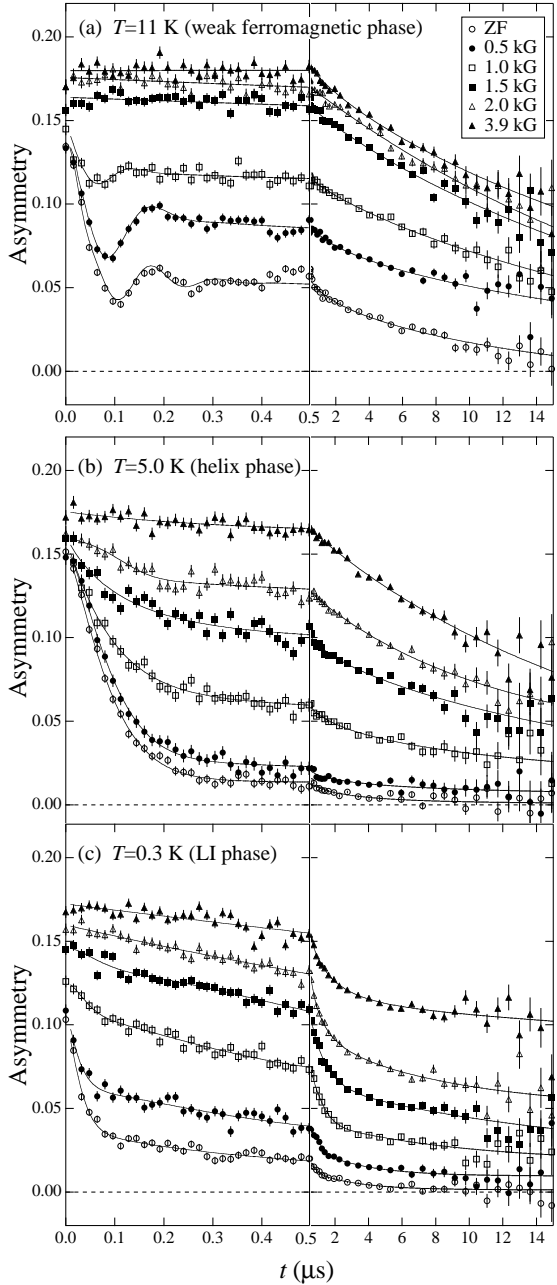}
\end{center}
\caption{Longitudinal field dependence of the time spectrum at (a) 11~K, (b) 5.0~K, and (c) 0.3~K. The left part of each panel shows the time spectra in the early-time region, $t<0.5~\mu$s, and the right one shows the time spectra in the long-time region, $t=0.5$$-$15~$\mu$s. The solid lines are guides to the eye.}
\end{figure}

As mentioned above, the dynamical nature of the spins is suggested in the ZF-$\mu$SR measurement. However, in order to discriminate the dynamic component from the static one in the local fields, we measured LF dependence of the time spectrum. In Fig.~2, the LF dependences of the time spectrum at 11~K, 5~K, and 0.3~K are shown as representative 
temperatures of the weak ferromagnetic, helix, and LI phases, respectively. For clarity, we show the time spectra in both time regions separately.

 At 11~K, the oscillations in the early-time region of $t< 0.5~\mu$s are observed in longitudinal fields, $H_{\rm LF} \leq 1$~kG. This is characteristic of static local fields. At the same time, the long-time tail ($t> \sim 0.2~{\rm \mu}$s) relaxes in both ZF and LF. Furthermore, the relaxation rate is almost independent of LF, suggesting that the characteristic time of the fluctuating component is rather short. Therefore, the LF dependence also shows the dynamical nature of local fields. At 5.0~K, the dynamical behavior of the local fields is more distinct. No oscillation is observed in both ZF and LF. At $t=0.2$$-$0.5~$\mu$s, the so-called decoupling behavior by LF is weaker than that at 11 K, though the relaxation at $t< 0.1~$$\mu$s is slower. The relaxation rate of the long-time tail decreases with LF, suggesting that the characteristic time of the fluctuating component becomes longer. On the other hand, the slightly round line shape at a fairly early time ($t < 0.05~\mu$s) and the decoupling behavior of the long time tail is consistent with the existence of the static component.
At 0.3~K, the feature of the LF dependence is similar to that at 5~K, although the line shape of the time spectrum is different. The decoupling behavior by LF is slightly strong than 5.0 K, which implies that the characteristic time of the fluctuating component becomes further longer.

\begin{figure}[t]
\begin{center}
\includegraphics[width=7.5cm]{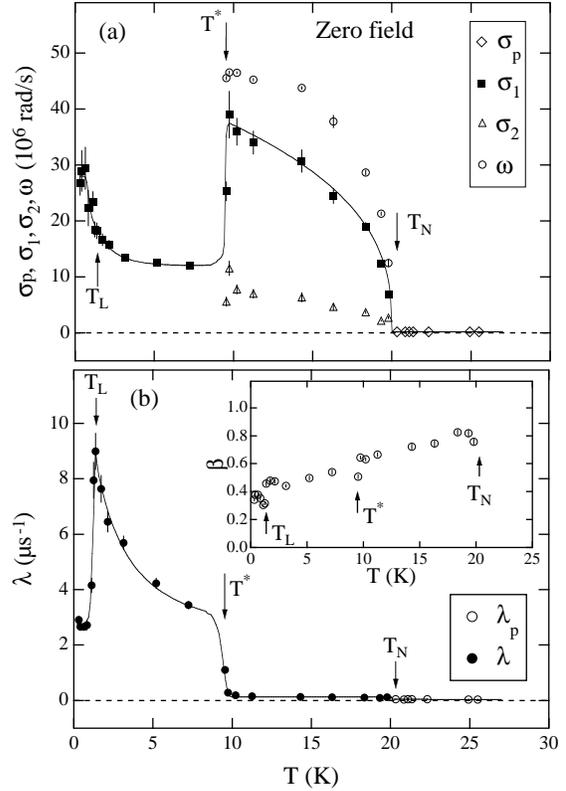}
\end{center}
\caption{Temperature dependence of (a) $\sigma_{\rm p}$, $\sigma_1$, $\sigma_2$, $\omega$, and (b) $\lambda$. The inset in (b) shows the temperature dependence of  
$\beta$. $T_{\rm N}$ and $T^*$ determined by the $M-T$ curve in $H = 100$~G (Fig. 1(a)) are shown by the arrows. $T_{\rm L}$ is defined as the temperature at which $\lambda$ shows a peak in this $\mu$SR measurement. The solid lines are guides to the eye.} 
\end{figure}


We attempted to find relaxation function which reproduce the time spectrum in both ZF and LF. However, we have not succeeded because of difficulty in estimation of the amplitude distribution of static component and the change of spin structure by applied magnetic fields. To qualitatively grasp the temperature dependence of the static and dynamic component, we fitted the time spectra in only ZF by phenomenological function as follows. In CuB$_2$O$_4$,  there are two Cu sites; Cu(A) spin which is almost static and Cu(B) spin which fluctuates even at 2~K.~\cite{Boehm03}
Therefore, we assumed that there exists no correlation between the static and dynamical components of the local fields. In this case, the relaxation function is given by the product of the static and dynamical relaxation 
functions,
\begin{equation}
G(t) = A_0~G_{\rm stat}(t) \times G_{\rm dyn}(t),
\end{equation}
where $A_0$ is the initial asymmetry of the time spectrum. 

In the P phase, the time spectra was fitted by using
$G_{\rm stat}(t) = \exp{(-\sigma_{\rm p}^2 t^2)}$ and 
$G_{\rm dyn}(t) = \exp{(-\lambda_{\rm p} t)}$.
The $G_{\rm stat}(t)$ approximates relaxation due to the nuclear dipole moments.

Below $T_{\rm N}$, the following phenomenological form is used for the static relaxation function,
\begin{eqnarray}
G_{\rm stat}(t) &=&  x_1 \exp{(-\sigma_1^2 t^2)}    \nonumber\\
 &&+ x_2 \exp{(-\sigma_2^2 t^2/2)} \cos{(\omega t +\phi)} +1/3.
\end{eqnarray}
The first, second, and third terms represent the relaxation at the early time ($t< \sim$0.2~$\mu$s), the oscillating component, and a powder-averaged constant value for the long-time tail, respectively. $x_1$ and $x_2$ denote the fractional amplitudes, and $\sigma_1$ and $\sigma_2$ represent the Gaussian relaxation rates of each term. The Larmor frequency $\omega$ and the Gaussian relaxation rates $\sigma_1$ and $\sigma_2$ reflect the static local fields at the muon site; the magnitude of the local fields and the distribution of the magnitude. $\phi$ is a phase offset caused by instrumental and sample-shape conditions.  Below 9.5~K, the second term was neglected, since no oscillation was observed. For the dynamic relaxation function, we used the following phenomenological form, a stretched exponential function,
\begin{equation}
G_{\rm dyn}(t) = \exp{(-(\lambda t)^\beta)},
\end{equation}
where $\lambda$ and $\beta$ represent the relaxation rate and power, respectively. 

The obtained parameters for the static component are 
summarized in Fig.~3(a). As mentioned above, 
below 9.5~K we cannot define parameters from the second term in eq.(2). As the temperature decreases, $\sigma_1$, $\sigma_2$, and $\omega$ begin to increase at $T_{\rm N}$.  Both $\omega$ and $\sigma_1$ show a conventional order parameter-like temperature dependence in the weak ferromagnetic phase. However, upon cooling, $\sigma_1$ abruptly decreases below $\sim$$T^*$. This decrease in the value of the static component is considered to be due to a change in the magnetic structure or in the dynamic behavior. Upon more cooling, $\sigma_1$ gradually increases again and saturates below $\sim$0.7~K. This temperature dependence of $\sigma_1$ is qualitatively consistent with the NMR line width measured in a magnetic field of $\sim$0.5~T~\cite{Nakamura04}, though the magnetic field affects the spin structure and the transition temperatures.

The parameters for dynamical components, $\lambda$ and $\beta$, are summarized in Fig. 3(b). A peak of $\lambda$, which generally suggests a second order phase transition, is not observed at $T_{\rm N}$. Upon cooling, $\lambda$ shows a step-like increase at $T^*$. A $\lambda$-type peak is observed at 1.4~K, suggesting a second order phase transition to the LI phase. Interestingly, below 0.7~K, $\lambda$ is almost constant, suggesting that the fluctuating component persists even at $T$$\rightarrow$0. On the other hand, the $\beta$ monotonically decreases below $T_{\rm N}$, suggesting a broadening of the distribution of the characteristic time of the spin fluctuations or the instantaneous amplitude of fluctuating fields. We also examined temperature dependence of $\lambda$ in $H_{\rm LF}$=0.35 T. Quite small peaks of $\lambda$ are observed at $T_{\rm N}$ and $T^*$ in addition to $T_{\rm L}$, and $T^*$ and $T_{\rm L}$ are lowered by magnetic field. The peak of $\lambda$ at $T^*$ is consitent with the NMR result under $\sim$0.5 T.~\cite{Nakamura05} 


In conventional magnets with a long-range order, the local fields are static in the ordered phase. However, in CuB$_2$O$_4$, the local fields are quasi-static in the weak ferromagnetic phase, whereas the fluctuating component becomes dominant at lower temperatures and persists even at 0.3~K. Such a thermal evolution is rather anomalous. Finally, we discuss two possible origins of the fluctuating fields observed below $T_{\rm N}$. 

One possible origin is the frustration of exchange interactions. 
 Although exchange interactions have not yet been determined in CuB$_2$O$_4$, the frustration is suggested from reduction of the Cu(B) moments at $T<2$~K.~\cite{Boehm03} The Cu(B) spins should fluctuate even below $T_{\rm N}$. 
In the weak ferromagnetic phase, the characteristic time of the fluctuations of the Cu(B) spins may be so short that the fluctuating component cannot strongly contribute to the $\mu$SR time spectra. Below $T^*$, the characteristic time becomes longer, and the fluctuating component may contribute more strongly to the time spectra.

As another possible origin of the fluctuations below $T^*$, we speculate the dynamic behavior of the soliton lattice. The formation of the soliton lattice, which is suggested to be a modulated helix structure, is proposed near $T^*$ by observing higher-order magnetic satellites in the neutron scattering measurements~\cite{Roessli01,Petrakovskii02}. However, the modulated helix structure may remain even below $T^*$. For example, if the propagation vector of domains and that of domain walls are close, the intensity of the higher-order satellites would be rather small and may not be observed.
In addition, in Ba$_2$CuGe$_2$O$_7$~\cite{Zheludev99}, Dzyaloshinskii-Moriya interactions and a special magnetic anisotropy, Kaplan-Shekht-Entin-Wohlman-Aharony term, form a soliton lattice in zero field. This special anisotropy always accompanies with Dzyaloshinskii-Moriya interactions in insulators~\cite{Zheludev99}. Therefore, we consider that the soliton lattice can be formed even below $T^*$, whereas no higher-order satellite was observed. Dynamic behaviors of the soliton lattice are reported in non-magnetic systems such as a charged soliton lattice system polyacetylene~\cite{Mori96} and a dielectric system Rb$_2$ZnCl$_4$~\cite{Hacker96}. Therefore, we consider that the fluctuating component observed in this $\mu$SR measurement might reflect the dynamic behavior of the magnetic soliton lattice like translational motion or fluctuations. If the dynamic behavior observed in this study is an essential phenomenon in a magnetic soliton lattice, it would be rather interesting. 

This work has been partly supported by the Hayashi Memorial Foundation for 
Female Natural Scientists.


\end{document}